\documentclass[fleqn,usenatbib]{mnras}

\usepackage{newtxtext,newtxmath}

\usepackage[T1]{fontenc}

\DeclareRobustCommand{\VAN}[3]{#2}
\let\VANthebibliography\thebibliography
\def\thebibliography{\DeclareRobustCommand{\VAN}[3]{##3}\VANthebibliography}

\usepackage{graphicx}	
\usepackage{amsmath}	

\title[Black hole distance]{On measuring the Hubble constant with X-ray reverberation mapping of active galactic nuclei}

\author[A. R. Ingram et al.]{Adam Ingram$^{1,2}$\thanks{E-mail: adam.ingram@newcastle.ac.uk},
Guglielmo Mastroserio$^{3}$,
Michiel van der Klis$^{4}$,
Edward Nathan$^{1}$,
Riley Connors$^{3}$,
\newauthor Thomas Dauser$^{5}$,
Javier A. Garc{\'\i}a$^{3}$,
Erin Kara$^{6}$,
Ole K\"{o}nig$^{5}$,
Matteo Lucchini$^{6}$
and Jingyi Wang$^6$
\\
$^{1}$Department of Physics, Astrophysics, University of Oxford, Denys Wilkinson Building, Keble Road, Oxford OX1 3RH, UK\\
$^{2}$School of Mathematics, Statistics and Physics, Newcastle University, Herschel Building, Newcastle upon Tyne, NE1 7RU, UK\\
$^{3}$Cahill Center for Astronomy and Astrophysics, California Institute of Technology, Pasadena, CA 91125, USA\\ 
$^{4}$Astronomical Institute, Anton Pannekoek, University of Amsterdam, Science Park 904, NL-1098 XH Amsterdam, Netherlands\\
$^{5}$Dr Karl Remeis-Observatory and Erlangen Centre for Astroparticle Physics, Universit\"{a}t
Erlangen-N\"{u}rnberg, Sternwartstr.~7, D-96049 Bamberg, Germany\\ 
$^{6}$MIT Kavli Institute for Astrophysics and Space
Research, MIT, 70 Vassar Street, Cambridge, MA 02139, USA\\
}

\date{Accepted 2021 October 8. Received 2021 September 15; in original form 2021 July 9}

\pubyear{2021}

\begin{document}
\label{firstpage}
\pagerange{\pageref{firstpage}--\pageref{lastpage}}
\maketitle

\begin{abstract}
We show that X-ray reverberation mapping can be used to measure the distance to type 1 active galactic nuclei (AGNs). This is because X-ray photons originally emitted from the `corona' close to the black hole irradiate the accretion disc and are re-emitted with a characteristic `reflection' spectrum that includes a prominent $\sim 6.4$ keV iron emission line. The shape of the reflection spectrum depends on the irradiating flux, and the light-crossing delay between continuum photons observed directly from the corona and the reflected photons constrains the size of the disc. Simultaneously modelling the X-ray spectrum and the time delays between photons of different energies therefore constrains the intrinsic reflected luminosity, and the distance follows from the observed reflected flux. Alternatively, the distance can be measured from the X-ray spectrum alone if the black hole mass is known. We develop a new model of our \textsc{reltrans} X-ray reverberation mapping package, called \textsc{rtdist}, that has distance as a model parameter. We simulate a synthetic observation that we fit with our new model, and find that this technique applied to a sample of $\sim 25$ AGNs can be used to measure the Hubble constant with a $3 \sigma$ statistical uncertainty of $\sim 6~{\rm km}~{\rm s}^{-1}{\rm Mpc}^{-1}$. Since the technique is completely independent of the traditional distance ladder and the cosmic microwave background radiation, it has the potential to address the current tension between them. We discuss sources of modelling uncertainty, and how they can be addressed in the near future.
\end{abstract}




\section{Introduction}

The Hubble constant, $H_0$, quantifies the current expansion rate of the Universe. The current best direct measurement via the Cepheid variable plus type 1a supernova distance ladder ($H_0 = 73.2 \pm 1.3~{\rm km}~{\rm s}^{-1}~{\rm Mpc}^{-1}$; \citealt{Riess2021}) is in $\sim 4.2 \sigma$ tension with the value inferred from fitting the power spectrum of cosmic microwave background (CMB) temperature anisotropies with the standard cosmological model ($H_0 = 66.5^{+1.3}_{-0.6}~{\rm km}~{\rm s}^{-1}~{\rm Mpc}^{-1}$ when neutrino mass density is free to vary; \citealt{Planck2020}). This discrepancy could indicate either that there are underestimated systematic uncertainties in the distance ladder \citep[e.g.][]{Efstathiou2014,Rigault2015,Follin2018,Bengaly2019,Riess2020,Mortsell2021}, or, more excitingly, that the standard cosmological model is incomplete \citep[e.g.][]{Knox2020,Vagnozzi2020}. New ways of measuring $H_0$ that are independent of the central two methods provide the possibility to differentiate between these two alternatives. 


There have already been many such attempts. Constraints from large scale structure surveys favour the CMB value \citep{Feeney2019,Philcox2020,DES2021}, whereas alternative direct measurement techniques have so far provided mixed results. Some studies using a re-calibrated distance ladder without Cepheid variables have supported the larger $H_0$ value \citep{Yuan2019,Huang2020}, and others have returned an intermediate value \citep{Freedman2019,Freedman2021}. A similar story applies to constraints from strongly lensed quasars \citep{H0LiCOW2020,Birrer2020} and spatially resolved nearby megamaser galaxies \citep{Pesce2020,Boruah2020}. Gravitational wave sources with electromagnetic counterparts can provide precise and independent $H_0$ constraints in the future \citep{Schutz1986,Nissanke2013,AbbotH02017} as long as enough events can be associated with a host galaxy \citep{Feeney2020}. Fast radio bursts have also been suggested as tools for cosmology \citep{WuFRB2020,Hagstotz2021}.

Here we consider active galactic nuclei (AGNs). Powered by accretion of material from the host galaxy onto the central supermassive black hole (BH), they are the most persistently luminous objects in the Universe and are therefore potentially useful for cosmology. Recent attempts to convert quasars to standardizable candles using the empirically derived non-linear relationship between X-ray and UV luminosity have yielded promising results \citep{Risaliti2015,Risaliti2019}.
However, the $L_\mathrm{UV}-L_\mathrm{X}$ relation cannot be used for independent measurement of $H_0$ because it is calibrated by the local distance ladder. For an independent measurement, a physical model for the spectrum is required. The central BH accretes in the form of a disc and, close to the BH itself, a cloud of hot electrons forms that is referred to as the \textit{corona}. The disc emits a multi-temperature blackbody spectrum in the optical and UV \citep{Shakura1973,Novikov1973}. Modelling the disc spectrum returns the luminosity as a function of disc size, which can in principle be constrained from light travel distances of the observed UV-optical time lags \citep{Collier1999,Cackett2007}. However, the time lags were often measured to be longer than expected for a standard disc model (the so-called \textit{disc size problem}), and consequently the inferred distances were much larger than expected. This problem can perhaps be solved by more realistic modelling of the time lag than was typically adopted in early studies \citep{Kammoun2021}.

Our proposed method instead makes use of the X-ray emission. The X-rays are dominated by the corona, which emits a power law spectrum due to Compton up-scattering of comparatively cool seed photons by hot coronal electrons \citep[e.g.][]{Thorne1975,Sunyaev1979}. In the case of the brightest AGNs, the corona is relatively compact ($\lesssim 20~R_g$ across, where $R_g=GM/c^2$ is a gravitational radius and $M$ is the BH mass; e.g. \citealt{Morgan2012}), but its exact geometry and physical origin is still an area of active research. Suggested models include a standing shock at the base of the jet (\citealt{Miyamoto1991,Fender1999,Markoff2005,Dauser2013})
and evaporation of the inner disk regions to form a hot, large scale height accretion flow (\textit{the truncated disk model}; \citealt{Eardley1975,Ichimaru1977,Done2007}).

Regardless of the origin of the corona, a fraction of the emitted photons irradiate the disc, where they are reprocessed and re-emitted with a characteristic \textit{reflection} spectrum. This reflection spectrum includes features such as an iron K-shell complex of emission lines at $\sim 6.4-7.1$ keV, an iron absorption K-edge at $\sim 7-9$\,keV, a broad profile peaking at $\sim 20-30$ keV referred to as the Compton hump (produced by electron scattering), and a soft X-ray excess \citep{George1991,Ross2005,Garcia2013}. The total observed spectrum is composed of (at least) the sum of two components: the power-law continuum of photons that travelled directly from the corona to the observer, and the reflection spectrum. The shape of the reflection spectrum depends on the properties of the illuminating spectrum and of the irradiated disc. Specifically, the shape and bolometric flux of the illuminating spectrum are both relevant, and the latter is typically specified by the illuminating X-ray flux, $F_x$. As for the disc, the relevant quantity is the density, $n_e$. Reflection
models typically take as input the ionisation parameter $\xi = 4\pi F_x / n_e$ \citep{Ross2005,Garcia2010}. For many years, $\xi$ was left as a free parameter in spectral fits, whereas disc density was hardwired to
$n_e = 10^{15}$ electrons per cm$^3$ \citep[e.g.][]{Walton2013,Garcia2015}, which is approximately expected for a BH mass of $M=10^8~M_\odot$ \citep{Shakura1973}. However, it has since been emphasized that the shape of the spectrum does not only depend on $\xi$, but also on $n_e$ itself \citep{Garcia2016}.
Newer models have therefore begun to include $n_e$ as a free parameter. Recent studies have shown that both $\xi$ and $n_e$ can be constrained by reflection spectroscopy, both for AGN \citep{Jiang2019AGN,Garcia2019}, and for their stellar-mass BH analogues, X-ray binaries \citep{Tomsick2018,Jiang2019GX339}. This means that modern X-ray reflection models can constrain $F_x$. The disc is highly optically thick and so all incident flux is re-emitted. The re-emission is multi-wavelength (i.e. not limited to X-rays) and is also not isotropic, but the full broad-band reprocessed spectrum is calculated as a function of emission angle
by modern reflection models \citep{Garcia2013}.
Therefore, for given values of $n_e$, $\xi$ and the inclination angle $i$, the locally re-emitted reflected flux is uniquely predicted by the reflection model. If the disc size is known, this reflected flux uniquely predicts the locally re-emitted reflected luminosity. The distance $D$ from Earth to the AGN follows from the comparison of the intrinsic reflected luminosity to the observed reflected flux crossing our X-ray detector.

The reflection spectrum is distorted by rapid orbital disc motion and gravitational redshift \citep{Fabian1989}, providing a diagnostic of the disc inclination angle and size. However, the relativistic photon energy shifts, and therefore the spectral distortions, are only sensitive to
size scales in units of $R_g$. X-ray spectroscopy alone can therefore be used to measure $M/D$, but mass and distance are completely degenerate. A number of recent studies have used known masses and distances for X-ray binaries to check that measured values of $\xi$ and $n_e$ are consistent with the observed reflected flux \citep{Shreeram2020,Zdziarski2020,Connors2021}.

Here, for the first time, we consider the prospects of using the relation between $\xi$, $n_e$ and observed flux to \textit{measure} distance, which requires the $M/D$ degeneracy to be broken. We suggest two ways to achieve this: either
inputting an independent mass measurement as a Bayesian prior to the spectral model
and/or
additionally modelling the light-crossing delay between variations in the direct and reflected spectral components (\textit{X-ray reverberation mapping}: \citealt{Campana1995,Reynolds1999,Uttley2014}). Since the time delays diagnose sizes in physical units, whereas spectral distortions are sensitive to the same size scales in units of $R_g$, modelling both together calibrates the length of a gravitational radius, and therefore constrains $M$. The light-crossing delay between continuum and reflected photons gives rise to reverberation features in a plot of time delay versus photon energy (a \textit{lag-energy spectrum}), such as a soft excess, iron line and Compton hump. Fourier frequency-dependent time lags between energy channels can be calculated from the argument of the cross-spectrum between the light curves from each of those channels \citep{vanderKlis1987}. A lag-energy spectrum can be built up from multiple cross-spectra, all of which feature one common \textit{reference band} light curve. A standard observational picture has emerged: at low Fourier frequencies ($\nu \lesssim 3 \times 10^{-4}~[10^6M_\odot/M]$ Hz) hard photons lag soft photons, both for the binaries \citep{Miyamoto1988,Nowak1999,Kotov2001} and AGNs \citep{Papadakis2001,McHardy2004,Epitropakis2017}. These lags are far larger than the expected light travel delay and are featureless in their energy dependence, and so are likely caused by spectral variability of the direct component rather than reverberation (possibly due to inward propagation of accretion rate fluctuations; \citealt{Arevalo2006,Ingram2013,Rapisarda2017a,Mahmoud2018}). For this reason, they are referred to as \textit{continuum lags}. At low frequencies, the continuum lag is so much larger than the expected reverberation lag that reverberation features are not unambiguously detectable. However, the magnitude of the continuum lag reduces with Fourier frequency, meaning that reverberation features can be detected at high frequencies. The first discoveries were in the form of the soft excess lagging the direct radiation (AGNs: \citealt{Fabian2009,Zoghbi2010,DeMarco2013}; X-ray binaries: \citealt{Uttley2011,DeMarco2015}), and later in the form of the iron line lagging the direct radiation (AGNs: \citealt{Zoghbi2012,Kara2016}; X-ray binaries: \citealt{Kara2019}).

A number of models have been developed for X-ray reverberation mapping \citep[e.g.][]{Cackett2014,Emmanoulopoulos2014,Epitropakis2016,Chainakun2016,Wilkins2016,Caballero-Garcia2018}. Our publicly available model \textsc{reltrans} \citep[][ hereafter I19]{Ingram2019} calculates the spectrum and lag-energy spectrum accounting for all relativistic effects, and reproduces continuum lags by introducing fluctuations in the power-law index of the spectrum radiated by the corona \citep{Mastroserio2018}. The effect that these power-law index fluctuations have on the reflection spectrum is self-consistently accounted for using a first order Taylor expansion. The model has already returned a BH mass of  $M\sim 25~M_\odot$ for the BH in Cygnus X-1, which agrees remarkably well with the recently updated dynamical measurement of $M = 22 \pm 2~M_\odot$; \citep{Miller-Jones2021}, and of $M \sim 10^6~M_\odot$ for Mrk 335, which is somewhat lower than the optical broad line region (BLR) reverberation mapping values of $M = [14.2 \pm 3.7] \times 10^6~M_\odot$ \citep{Peterson2004} and $M = [26 \pm 8]\times 10^6 ~ M_\odot$ \citep{Grier2012}. These results all used the original version of our model, which calculated the restframe reflection spectrum (using the model \textsc{xillver}: \citealt{Garcia2013}) with $n_e=10^{15}~{\rm cm}^{-3}$ hardwired. However, after subsequent updates $n_e$ is a free parameter in the most recent version of the model \citep[version 2.0:][]{Mastroserio2021,Wang2021}.
Here, we further update the model such that $D$ is now a free parameter instead of $\xi$, which is instead self-consistently calculated. In Section \ref{sec:model}, we describe this update to the model alongside some other new features that give the model more flexibility to explore systematics associated with the uncertain coronal geometry. In Section \ref{sec:sims}, we use the new model to simulate a bright AGN and fit back to the synthetic data in order to determine the statistical precision we can achieve on $H_0$ with a single AGN. In Section \ref{sec:discussion}, we calculate the sample size of AGN that would be required to reduce this to a level low enough to address the current tension, as well as discussing sources of systematic error. We conclude in Section \ref{sec:conclusions}.

\section{The model}
\label{sec:model}

We develop a new flavour of the \textsc{reltrans} model called \textsc{rtdist}. All \textsc{reltrans} models calculate direct continuum and reflection components assuming a small spherical corona located above the BH on its rotational axis (the \textit{lamppost model}), to output a time-averaged flux spectrum and an energy-dependent cross-spectrum for an input range of Fourier frequencies. We start from \textsc{reltrans} v2.0 \citep{Mastroserio2021} and add several new features. The key improvement is that distance is now a parameter in place of ionization parameter. The specific model parameter is angular diameter distance. The relevant definition of distance for calculating $H_0$ is proper distance, which is equal to angular diameter distance in a flat universe (as is assumed throughout this paper). We also include two new features designed to give the model more flexibility to explore systematics associated with the uncertain coronal geometry. Firstly, a flat disc in the BH equatorial plane was previously assumed, whereas we now generalise this to a conical disc with scale-height $h_d/r$ left as a model parameter. The old flat disc assumption can be recovered by setting $h_d/r=0$. Secondly, we include an angular emissivity profile of the corona instead of assuming isotropic emission. Both considerations influence the radial dependence of irradiating X-ray flux on the disc. Throughout, lengths denoted by lower case symbols are in units of $R_g$, whereas upper case symbols correspond to physical units (e.g. $r = R/R_g$).

\subsection{Including an arbitrary angular emissivity}

We assume that there is one visible corona -- a lamppost source at height $h~R_g$ above the BH. There is also presumably another similar source blocked from view beneath the disc. The specific intensity radiated by the visible source in its own restframe is
\begin{equation}
    I_s(E_s,t,\theta,\phi) = \frac{C(t)}{(a_s/4)} f(E_s) p(\theta,\phi).
    \label{eqn:Is}
\end{equation}
Here, $E_s$ is photon energy measured in the source restframe, $t$ is coordinate time and $a_s$ is the surface area of the source. Since the source is assumed to be small\footnote{Meaning that a locally Euclidean spacetime can be assumed.} and spherical, its area projected onto the image plane is simply $a_s/4$. The function $p(\theta,\phi)$ encapsulates the angular dependence and is normalised such that
\begin{equation}
    \int_0^{2\pi} \int_{-1}^{+1} p(\theta,\phi) ~d\mu~d\phi = 1,
    \label{eqn:intp}
\end{equation}
where $\mu \equiv \cos\theta$, and $\theta$ and $\phi$ are the polar and azimuthal angles in a right-handed coordinate system with z-axis aligned with the BH spin axis. Given this normalisation, it is straight forward to recover Equation (8) from I19. The energy dependence is encapsulated in the function $f(E_s)$ (in units of specific luminosity), which is calculated from the \textsc{xspec} model \textsc{nthComp} \citep{Zdziarski1996}. This produces a power law spectrum (specific photon flux $\propto E^{-\Gamma}$) with low and high energy cut offs set respectively by the seed photon temperature and electron temperature of the corona, $kT_e$. $C(t)$ is a dimensionless time-dependent normalisation constant.

It can be shown from following the derivation in I19 that the specific flux seen by a distant observer is
\begin{equation}
    F(E,t) = A(t) ~\ell ~g_{\rm so}^{\Gamma(t)}~f(E|kT_{\rm e,obs})~4\pi p(\theta_o,\phi_o),
    \label{eqn:Fo}
\end{equation}
where $A \equiv C / (4 \pi D^2)$, $D$ is angular diameter distance from the observer to the BH and $\ell$ (see Equation 11 of I19) accounts for gravitational lensing around the BH. Here, $g_{\rm so}=E/E_s$ is the energy shift experienced by photons when traveling from source to observer (also including the redshift due to the motion of the host galaxy, which is a combination of the Hubble flow and peculiar velocity) and $kT_{\rm e,obs} = g_{\rm so} kT_{\rm e}$.
The angles $\theta_o$ and $\phi_o$ specify the initial trajectory of photons that eventually hit the observer's detector (at polar angle $i$).

The combined luminosity of the two lamppost sources relates to $A$ as
\begin{equation}
    L_{\rm corona} = A~8\pi~D^2~g_{\rm so}^{\Gamma-2}~\int_0^\infty ~f(E)~dE.
    \label{eqn:Lacc}
\end{equation}
This is the accretion luminosity
in the limiting case whereby
the corona completely dominates the radiation output (i.e.
if the intrinsic disc emission is low). Since $L_{\rm corona}$ can be readily interpreted physically, it is tempting to make it a model parameter. However, we choose instead to use $A$ as the normalisation parameter for the continuum spectrum in our model, as this leads to a simpler model parameter space than using $L_{\rm corona}$ (or indeed $C$), therefore making it easier (thus potentially faster) for minimisation algorithms to find a global minimum in $\chi^2$. Clearly, the more physically interesting quantity $L_{\rm corona}$ can be simply calculated from the best fitting $A$ value.

Rays initially emitted from the corona with angles $\theta_d$ and $\phi_d$ hit the disc at radius $r$ and azimuth $\phi$. Irradiating photons are reprocessed in the disc atmosphere and are re-emitted. The specific flux of re-emitted photons seen by a distant observer from a disc patch with radial and azimuthal extent $dr$ and $d\phi$ is
\begin{equation}
    dR(E,t) = A(t-\tau(r,\phi))~g_{\rm do}^3(r,\phi)~\epsilon(r)~\mathcal{R}(E_d)~d\alpha~d\beta,
    \label{eqn:dRef}
\end{equation}
where $\alpha$ and $\beta$ (the impact parameters at infinity) are horizontal and vertical distances on the image plane, $\mathcal{R}(E_d)$ is the restframe reflection spectrum (calculated using \textsc{xillverDCp}; \citealt{Garcia2013,Garcia2016}), $g_{\rm do}$ is the energy shift experienced by photons travelling from disc patch to observer (see Appendix \ref{sec:gfac}), and $\tau$ is the time delay of reflected photons with respect to directly observed ones. The disc emissivity is
\begin{equation}
    \epsilon(r) = 2\pi~p(\theta_d,\phi_d)~g_{\rm sd}^\Gamma(r)~\frac{|d\cos\delta/dr|}{dA_{\rm ring}/dr},
    \label{eqn:emissivity}
\end{equation}
where $\delta = \pi - \theta_d$, $g_{\rm sd}$ is the energy shift experienced by photons as they propagate from the source to a radius $r$ on the disc (see Appendix \ref{sec:gfac}) and $dA_{\rm ring}/dr$ is as defined in I19 (equation A1 therein). Note that $\epsilon(r)$ is in units of $1/R_g^2$. The above equation trivially reduces to Equation (20) of I19 for $p=1/(4\pi)$ (corresponding to an isotropic source). We use the public general relativistic ray tracing code \textsc{ynogk} (\citealt{ynogk}; based on \textsc{geokerr}: \citealt{Dexter2009}) to map the angle $\theta_d$ to the disc radius $r$ and to map disc coordinates to impact parameters.

\subsection{Form of the angular emissivity}

\begin{figure}
\centering
\includegraphics[width=\columnwidth,height=1.05\columnwidth,trim=1.5cm 1.5cm 2.0cm 11.0cm,clip=true]{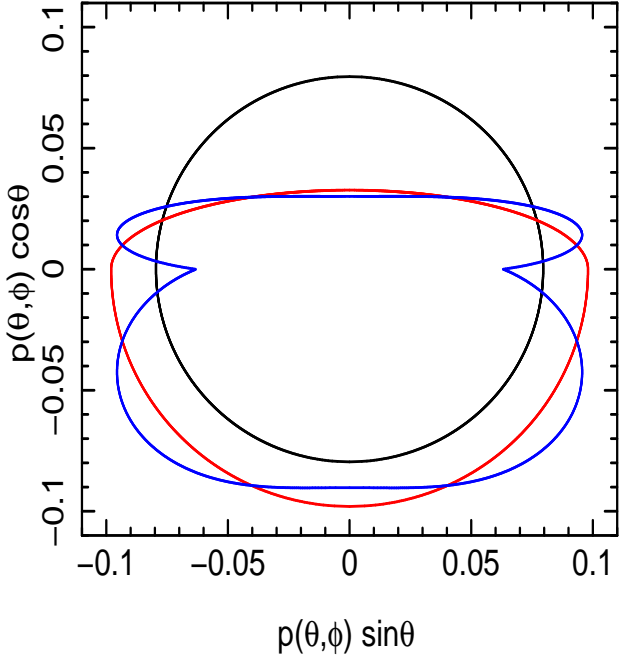}
\vspace{-5mm}
\caption{Visualisation of our angular emissivity profile $p(\theta,\phi)$, where $\theta$ is measured clockwise from the vertical. An isotropic emissivity profile (black) results from setting $b_1=b_2=0$, \texttt{boost}=1, whereas \texttt{boost}=3 (red) results in strong downwards boosting (i.e. boosting towards the disc). Setting $b_1=0.429$, $b_2=-2$, \texttt{boost}=3 (blue) produces a profile that is still focused downwards and has additional structure.}
\label{fig:pvis}
\end{figure}

\begin{figure}
\centering
\includegraphics[width=\columnwidth,trim=0.5cm 1.5cm 2.0cm 4.0cm,clip=true]{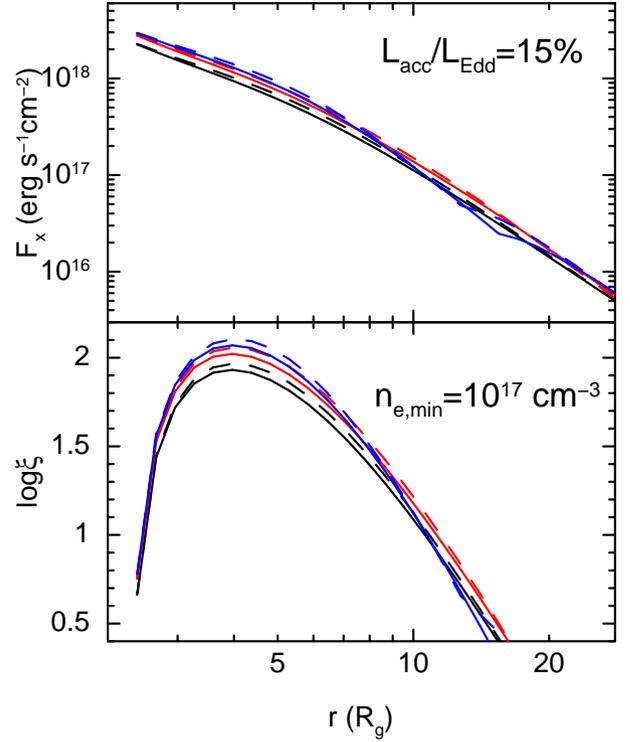}
\vspace{-5mm}
\caption{Irradiating flux (top) and ionization parameter (bottom) as a function of radius for model parameters: $h=6$, dimensionless BH spin parameter $a=0.9$, $r_{\rm in}$ is set to the innermost stable circular orbit (ISCO) $r_{\rm isco}(a)$,  $\Gamma=2$, $M=10^7~M_\odot$, $D=100$ Mpc, $A=10^{-4}$, $n_{\rm e,min}=10^{17} {\rm cm}^{-3}$, and we assume a `Zone A' density profile. The model normalisation is set such that the coronal luminosity (Equation \ref{eqn:Lacc}) is $15\%~L_{\rm Edd}$. Solid and dashed lines represent $h_d/r=0$ and $h_d/r=0.1$ respectively. Colours have the same meaning as in Fig. \ref{fig:pvis}: black is $b_1=b_2=0$, \texttt{boost}=1; red is $b_1=b_2=0$, \texttt{boost}=3; and blue is $b_1=0.429$, $b_2=-2$, \texttt{boost}=3. The dip in the blue lines at $r \sim 15$ corresponds to the dip in $p(\theta,\phi)$ at $\theta=\pi/2$ from Fig. \ref{fig:pvis}.}
\label{fig:logxir}
\end{figure}

We assume a quadratic dependence of $p$ on $\mu \equiv \cos\theta$ (following \citealt{Ingram2015a,Ingram2017})
\begin{equation}
    p(\theta,\phi) \propto 1 + ( b_1 + |b_2|) |\mu| + b_2 \mu^2,
    \label{eqn:psimple}
\end{equation}
where $b_1 \geq 0$ and $b_2$ are two dimensionless angular emissivity parameters. The choice of $b_1+|b_2|$ as the linear coefficient is to ensure that $p(\theta,\phi)$ never becomes negative for any value of $\theta$ as the $b_1$ and $b_2$ parameters are varied in a fit loop. We normalise the function such that condition (\ref{eqn:intp}) is satisfied. Although our formalism allows for an azimuthal dependence, we assume azimuthal symmetry throughout this paper.

The above profile is symmetric about $\mu=0$, where $\mu > 0$ corresponds to photons from the corona initially travelling away from the disc plane, and $\mu < 0$ to photons initially travelling towards the disc. However, there may in reality be boosting. For example, if the electrons in the corona are moving away from the BH, then radiation will be boosted away from the BH such that $p(\mu=1) > p(\mu=-1)$. In previous versions of \textsc{reltrans}, we accounted for this by simply multiplying the output reflection spectrum by the `boost' parameter, \texttt{boost}$=1/\mathcal{B}$, which is left as a model parameter. This treatment is not, however, entirely self-consistent because the disc emissivity $\epsilon(r)$ is calculated assuming isotropic radiation. For the new model, we instead re-purpose the boost parameter to adjust the angular emissivity profile such that \texttt{boost}$= p(\mu=-1) / p(\mu=1)$. The boost parameter therefore has the same basic functionality as before -- increasing it increases the amount of reflection in the spectrum relative to the continuum level -- but in the new treatment the value of the boost parameter also adjusts the radial form of $\epsilon(r)$.

The exact function we use is 
\begin{equation}
    p(\theta,\phi) \propto \left( 1 + ( b_1+|b_2| ) |\mu_p| + b_2 \mu_p^2 \right) \sqrt{1+\mu_p^2 (N_\pm-1)},
    \label{eqn:pform}
\end{equation}
where
\begin{equation}
    \mu_p = \left[ N_\pm^2 (\mu^{-2}-1) + 1 \right]^{-1/2},
\end{equation}
and $N_\pm=\mathcal{B}$ for $\mu>1$ and $N_\pm=1$ for $\mu<1$. This expression reduces to Equation (\ref{eqn:psimple}) when \texttt{boost}=1. Fig. \ref{fig:pvis} shows visualisations of the angular emissivity profile for three parameter combinations. Isotropic radiation ($b_1=b_2=0$, \texttt{boost}=1) is represented by a circle (black line), indicating that the intensity is the same in all directions. Adjusting to \texttt{boost=3} (still with $b_1=b_2=0$) leads to radiation being beamed downwards towards the BH (red line). The blue line represents non-isotropic radiation with $b_1=0.429$, $b_2=-2$ and \texttt{boost}=3. This combination of $b_1$ and $b_2$ corresponds to the intensity from a Comptonising slab with optical depth $\tau=1$ \citep{Sunyaev1985,Ingram2015a}. Note that we do not expect a spherical source to radiate like a Comptonising slab; we are simply using this parameter combination to illustrate the function.

We note that the distance measured by fitting the model to data is at most only very weakly dependent on the assumed form of $p(\theta,\phi)$ (see Section \ref{sec:results}). This is because the illuminating flux $F_x$ is related directly to the integrated flux in the observed reflection spectrum. There is therefore no strong degeneracy between $p$ and $F_x$. The form of $p(\theta,\phi)$ instead influences the form of $\epsilon(r)$, which in turn subtly influences the shape of the iron line profile in the observer frame.

\subsection{Ionization parameter and distance}

The ionization parameter is, in general, a function of disc radius: $\xi(r) = 4\pi F_x(r) / n_e(r)$. Whereas the old \textsc{reltrans} model included the maximum value of the radial function $\log\xi(r)$ as a model parameter, the new model instead calculates $\xi(r)$ self-consistently from the following
\begin{equation}
    \xi(r) = \left(\frac{4\pi c^2 D}{G M}\right)^2 A \frac{\epsilon(r)}{n_e(r)}~\left(\frac{g_{\rm sd}(r)}{g_{\rm so}}\right)^{2-\Gamma}\int_{0.1~{\rm keV}}^{1~{\rm MeV}} f(E) dE,
    \label{eqn:xi}
\end{equation}
where $A$ is the time-averaged value of $A(t)$ and $D$, $A$ and $M$ are model parameters\footnote{Also note that, in line with the definitions in the newest version of \textsc{xillverDCp}, $F_x(r)$ is defined as the illuminating flux in the photon energy range $0.1~{\rm keV} \leq E \leq 1~{\rm MeV}$. This range differs from the definition in older \textsc{xillver} models. The new energy range is consistent with \textsc{reflionx} \citep{Ross2005}. Note that the integration range used is not important as long as it is treated consistently, since the entire broad-band illuminating spectrum is specified within the model. Therefore $F_x$ simply serves as a normalisation of the illuminating spectrum within the model.}. As before, our model includes two different forms for $n_e(r)$: either a constant or the `zone A' density profile (see e.g. Table 2 of \citealt{Shreeram2020}) from the \citet{Shakura1973} disc model. The density profile is normalised by the model parameter $\log n_{\rm e,min}$, which is the minimum value of the radial function $\log n_e(r)$. It is clear from Equation (\ref{eqn:xi}) that the X-ray spectrum is sensitive to the ratio $D/M$. Additionally considering timing information provides orthogonal constraints on $M$, which in turn allows $D$ to be constrained.


Fig. \ref{fig:logxir} shows example radial profiles of irradiating flux (top) and ionization parameter (bottom) for $M=10^7~M_\odot$, $A=10^{-4}~{\rm cm}^{-2}$ and $D=100$ Mpc. These parameters correspond to the BH accreting at $15\%$ of its Eddington limit. Different colours correspond to different coronal angular emissivity profiles, and we employ the same colour coding as Fig. \ref{fig:pvis}. Solid lines are for $h_d/r=0$ and dashed lines are for $h_d/r=0.1$, which is already a reasonably thick disc. We see that, consistent with previous studies of the emissivity in the lamppost model \citep[e.g.][]{Wilkins2012,Dauser2013,Ingram2019}, the irradiating flux is $F_x \propto r^{-3}$ for $r \gtrsim h = 6$, and is a flatter function of radius for $r \lesssim h$. This is reasonably consistent with the Newtonian expectation, since the strongest relativistic effects are only important for lower source height and higher BH spin \citep[e.g.][]{Ingram2019}. We also see that the angular emissivity parameters $b_1$ and $b_2$ have only a subtle effect on the profile, with the flux for the blue line being slightly higher for low $r$ at the expense of the flux at higher $r$. The boost parameter has a more obvious effect, with larger \texttt{boost} leading to more overall irradiating flux, as well as a subtle change in the radial profile. The disc scale height also has only a subtle effect on the profile, with thicker discs intercepting slightly more flux from the corona.
In the bottom panel of Fig. \ref{fig:logxir}, we plot the ionization parameter corresponding to the above $F_x(r)$ function for a zone A density profile with $n_{\rm e,min}=10^{17}~{\rm cm}^{-3}$. The ionization parameter reaches its maximum value at approximately the radius at which the density reaches its minimum value ($r\sim 4$).


\subsection{Model outputs}

The rest of the model calculation is identical to that of \textsc{reltrans} v2.0 \citep{Mastroserio2021}. The total time dependent spectrum is the sum of direct and reflected components $S(E,t)=F(E,t)+R(E,t)$, where $F(E,t)$ is given by Equation (\ref{eqn:Fo}) and $R(E,t)$ is given by the integral of Equation (\ref{eqn:dRef}) over all impact parameter combinations for which $r_{\rm in} \leq r \leq r_{\rm out}$. The time-averaged flux spectrum is then simply $S(E,t)$ for the time-averaged values of $A(t)$ and $\Gamma(t)$. The cross-spectrum is $G(E,\nu) = S(E,\nu) F_r^*(\nu)$, where $F_r(\nu)$ is the Fourier transform of the time-dependent flux in the reference band, which is the count rate in an energy range specified by the user. The cross-spectrum is calculated by Taylor expanding $A(t)$ and $\Gamma(t)$ around their time-averaged values, leading to $G(E,\nu)$ being expressed as the sum of four transfer functions \citep[see][for details]{Ingram2019,Mastroserio2021}. The cross-spectrum $G(E,\nu)$ is then averaged over a user-defined frequency range (which should be the same range as was used to average the data) to get $\langle G(E,\nu_c)\rangle$ \citep[see][for details]{Mastroserio2020}. The fluctuations in $A(t)$ and $\Gamma(t)$ in a given frequency range are specified by the parameters $\phi_{AB}$ and $\gamma$. Here, $\phi_{AB}$ is the phase difference between fluctuations in $\Gamma(t)$ and those in $A(t)$ (positive means that $\Gamma$ lags $A$), and $\gamma$ is the variability amplitude of $\Gamma(t)$ relative to the fractional variability amplitude of $A(t)$ (see \citealt{Mastroserio2021} for details). The model can output the time-averaged flux spectrum and the cross-spectrum in multiple frequency ranges. The cross-spectrum can be output either as real and imaginary parts, or as time lag vs energy and absolute rms vs energy spectra. In this paper, we consider only the time-averaged flux spectrum and the lag-energy spectrum.

\section{Simulations}
\label{sec:sims}

Since our model is sensitive to distance, it is in principle possible to measure the Hubble constant, $H_0 = z c / D$, by modelling the spectrum and lag spectrum of bright AGNs ($z$ is the host galaxy redshift). In this Section, we run a simulation to investigate the statistical uncertainties we may expect to achieve for a typical exposure of a bright type 1 AGN. This statistical uncertainty can then be reduced by repeating the analysis for many sources. The analysis is of course also subject to many potential sources of systematic error, which we discuss in the following section.

\subsection{Method}

Our model outputs the flux spectrum and the energy-dependent cross-spectrum for a given Fourier frequency range, which contains within it information both on the correlated variability amplitude and the time lags between energy bands (the lag-energy spectrum). Here we consider only the flux spectrum and the lag-energy spectrum in various frequency ranges. We simulate the flux spectrum using the \textsc{xspec} routine \texttt{fakeit}.

In order to simulate lag-energy spectra, we develop our own \textsc{xspec} user-defined model. Time lags are constrained observationally by first extracting light curves for a number of energy bands (the \textit{subject} bands, $s(E_I,t)$) and in one broad energy band (the \textit{reference} band, $r(t)$), which is often the sum of all the subject bands\footnote{Note that it is prevalent in the literature to use as the reference band the sum of all bands minus the current subject band. This is to ensure that the subject and reference bands are statistically independent of one another. However, \citet{Ingram2019a} showed that there is no need to subtract the current subject band from the reference band. Statistical independence can be recovered simply by subtracting off a Poisson noise contribution to the cross-spectrum. The \citet{Ingram2019a} approach not only increases the signal to noise slightly by enabling a higher reference band count rate, but it also eliminates the bias introduced by using a different reference band for each subject band.} \citep[e.g.][]{Uttley2014,Ingram2019a}. A cross-spectrum, $\tilde{G}(E_I,\nu_c) = \langle S(E_I,\nu) R^*(\nu) \rangle$, is then calculated for each subject band, where $R(\nu)$ is the Fourier transform of $r(t)$, the angle brackets denote an average over multiple realisations, and the tilde denotes an observationally estimated quantity. The lag-energy spectrum for a frequency range centred on $\nu_c$ is $\tilde{t}_{\rm lag}(E_I,\nu_c)={\rm arg}[\tilde{G}(E_I,\nu_c)]/(2\pi\nu_c)$.  This averaging is typically a combination of averaging over a number of separate light curve segments (\textit{ensemble averaging}) and over adjacent (statistically independent) Fourier frequencies. These two processes are statistically equivalent, and the total number of realisations is simply the number of segments multiplied by the number of frequencies averaged over \citep{vanderKlis1989}. For averaging over a frequency range $\nu_{\rm min}$ to $\nu_{\rm max}$, regardless of the nature of ensemble averaging employed, the total number of realisations is $N=T(\nu_{\rm max}-\nu_{\rm min})$, where $T$ is the total exposure time \citep{Ingram2017a,Ingram2019a}. If $N\gtrsim 40$, the errors on the lag-energy spectrum are Gaussian to a good approximation \citep{Huppenkothen2018,Ingram2019a}.

We therefore simulate Gaussian random variables. The mean is the model output $t_{\rm lag}(E_I,\nu_c)=\varphi(E_I,\nu_c)/(2\pi \nu_c)$, where $\varphi(E_I,\nu_c)$ is the argument of the cross-spectrum $G(E_I,\nu_c)$. Note that $G(E_I,\nu_c)$ has been folded around the instrument response matrix\footnote{The mathematical operation is $G(E_I,\nu_c) = \int_0^\infty R_D(E_I,E) G(E,\nu_c) dE$, where $R_D(E_I,E)$ is the response function of the $I^{\rm th}$ energy channel and the telescope effective area for a photon with true energy $E$ is the sum of $R_D(E_I,E)$ over all energy channels. In practice, $R_D(E_I,E)$ is discretized into the response matrix, $R_D(E_I,E_J)$.}. The standard deviation is the $1\sigma$ error $dt_{\rm lag}(E_I,\nu_c)=d\varphi(E_I,\nu_c)/(2\pi \nu_c)$, which we calculate using an error formula derived from Equation (19) of \citet{Ingram2019a}, where
\begin{eqnarray}
    &&[d\varphi(E_I,\nu_c)]^2 = \frac{1+P_{\rm r,n}/P_r(\nu_c)}{2 T (\nu_{\rm max}-\nu_{\rm min})} \nonumber \\
    && \frac{(1-\gamma_c^2(\nu_c))|G(E_I,\nu_c)|^2 + P_{\rm s,n}(E_I) P_r(\nu_c)}{\gamma_c^2(\nu_c) |G(E_I,\nu_c)|^2},
\end{eqnarray}
and all symbols in the above that have not yet been introduced are defined in the rest of this sub-section. Our model implicitly assumes unity coherence between all energy bands, where coherence is a measure of how well one light curve is linearly correlated with another and takes a value between zero and unity \citep[e.g.][]{Vaughan1997,Nowak1999}. Because of this, the power spectrum of the reference and subject bands predicted by the model, $P_r(\nu_c)$ and $P_s(E_I,\nu_c)$, relate to the model cross-spectrum as 
$|G(E_I,\nu_c)|^2 = P_s(E_I) P_r(\nu_c)$. These power spectra are in the absence of Poisson noise. We calculate the expected Poisson noise levels of the reference and subject bands, $P_{r,n}$ and $P_{r,s}(E_I)$, from the predicted reference and subject band count rates, $\mu_r$ and $\mu_s$, as $P_{r,n}=2\mu_r$ and $P_{s,n}(E_I) = 2\mu_s(E_I)$ \citep[e.g.][]{vanderKlis1989,Uttley2014}. Note that this is employing absolute rms normalisation, such that integrating the power spectrum over all frequencies gives the variance of the corresponding light curve \citep{vanderKlis1989}. Since the reference band is assumed to be the count rate summed over energy channels $I_{\rm min} \leq I \leq I_{\rm max}$, the reference band power spectrum is equal to
\begin{equation}
    P_r(\nu_c) = \sum_{I=I_{\rm min}}^{I_{\rm max}} {\rm Re}[G(E_I,\nu_c)],
\end{equation}
and the subject band power spectrum is $P_s(E_I,\nu_c) = |G(E_I,\nu_c)|^2 / P_r(\nu_c)$. Since we do not model the modulus of the cross-spectrum in this analysis, only its argument, we must input additional information into our simulation about the variability amplitude. Namely, we specify the power spectrum of the spectral normalisation in fractional rms normalisation, $|A(\nu_c)|^2/A_0^2$, as a simulation input. In contrast to our model, the observed coherence between two X-ray bands tends to be unity only up to a high frequency break, $\sim 5 (10 M_\odot/M)$ Hz \citep[e.g.][]{Vaughan1997,Nowak1999,Uttley2014}. In order to account for this, we introduce a coherence $\gamma_c(\nu_c)$ to our simulations, such that $\tilde{G}(E_I,\nu_c) \approx \gamma_c(\nu_c) G(E_I,\nu_c)$, which increases the error estimate from what would be predicted from the model alone. This is also set as a simulation input. Since we only fit for the time lags, the fit is insensitive to $|A(\nu_c)|^2/A_0^2$ and $\gamma_c(\nu_c)$. These parameters only set the uncertainties on the simulated time lags. All other quantities required for the simulation  are already defined within the model.

\subsection{Set up}

We simulate an observation with $260$ ks \textit{XMM-Newton} exposure and $100$ ks 
exposure on each focal plane module (FPM) of \textit{NuSTAR}, basing our source parameters on Ark 564. This is a perfect target source, since an iron K reverberation feature has previously been detected in its lag-energy spectrum \citep{Kara2013,Kara2016}, and disc density constraints have also been placed from X-ray spectral fits \citep{Jiang2019AGN}. \textit{XMM-Newton} and \textit{NuSTAR} have both observed the source for a total of $>500$ ks \citep{Kara2016}, but we choose a shorter exposure to make our simulation more representative of other sources for which either the total archival exposure time is less, or the count rate is lower. We base the input spectral parameters on the model of \citet{Jiang2019AGN} (see their Fig. A1 and their Table A2; also see \citealt{Walton2013}). All of our input model parameters are listed in Table \ref{tab:fitback}. For simplicity, we fix the disc inner radius to the ISCO and set $a=0.9$. We choose an intermediate source height of $h=6$ and set the disc scale height to $h_d/r=0.02$. We assume an isotropically emitting source ($b_1=b_2=0$, boost=1), a Zone A radial density profile and set the minimum of this profile to $\log(n_e/{\rm cm}^{-3})=17$. This is smaller than the $\log(n_e/{\rm cm}^{-3})\approx 18.5$ density value inferred by \citet{Jiang2019AGN}, but that was assuming that density is independent of radius. We assume a distance of $100$ Mpc. For the redshift of $z=0.024917$ this corresponds to a Hubble constant of $H_0=74.70~{\rm km}~{\rm s}^{-1}~{\rm Mpc}^{-1}$, which is close to the current distance ladder measurement \citep[e.g.][]{Riess2021}. Consistent with Ark 564, the $1-10$ keV flux of the model is $3.55\times 10^{-11}~{\rm ergs}~{\rm cm}^{-2}~{\rm s}^{-1}$. We set the BH mass to 3 million $M_\odot$, for which the corona luminosity is $L_{\rm corona} \sim L_{\rm Edd}$. We note that the slightly lower BH mass of $M\sim[1-2]$ million $M_\odot$ often assumed in the literature \citep[e.g.][]{Kara2013,Jiang2019AGN} implies a super-Eddington coronal luminosity for the moderate source height we assume here. Although \citet{Jiang2019AGN} did not find a narrow reflection component (e.g. from the AGN torus) to be statistically required in their model of Ark 564, we include one here in the form of a \textsc{xillverCp} component to be conservative and to make our model more generally applicable to other AGN. Since it originates from a very large emitting region, we assume that it does not vary on short timescales. We therefore include it in the flux spectrum, but not in the lag-energy spectrum.

\begin{figure}
\centering
\includegraphics[width=\columnwidth,trim=0.5cm 1.5cm 2.5cm 2.0cm,clip=true]{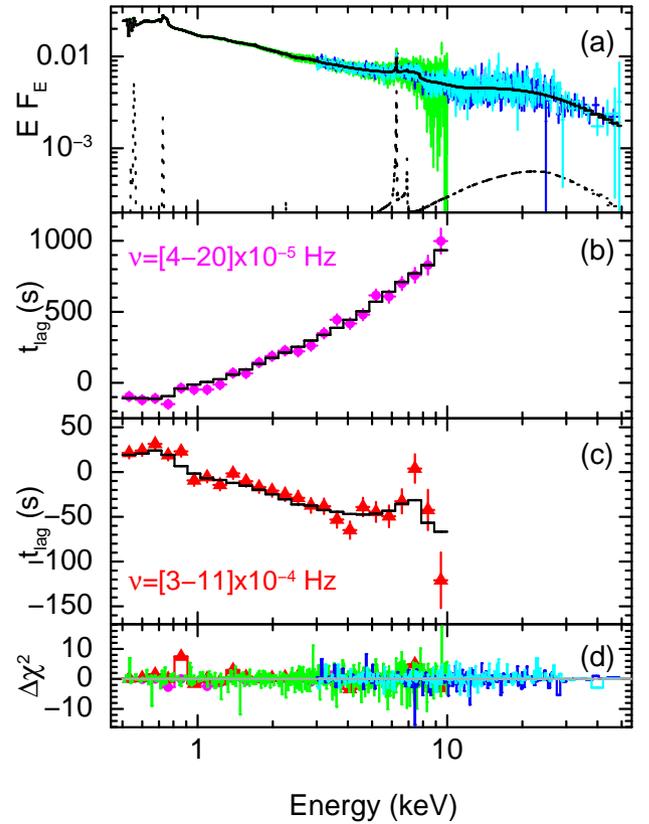}
\vspace{-5mm}
\caption{Synthetic data ($260$ ks \textit{XMM-Newton} EPIC-pn, $100$ ks \textit{NuSTAR}) and Model 1. \textit{(a):} Flux spectrum (${\rm keV}~{\rm cm}^{-2}~{\rm s}^{-1}$) of fake \textit{XMM-Newton} (green), FPMA (blue) and FPMB (cyan) data
unfolded around the best fitting model (black solid line). The dotted line is the distant reflector. \textit{(b-c):} Lag-energy spectrum in Fourier frequency ranges as labelled. \textit{(d):} Contributions to $\chi^2$, with colours corresponding to the above panels. Reduced $\chi^2$ is $\chi^2/{\rm d.o.f.}=2525.4/2606$ (null-hypothesis probability $= 0.868$).}
\label{fig:fitback}
\end{figure}

We simulate flux spectra for both \textit{XMM-Newton} and \textit{NuSTAR}, and lag-energy spectra for \textit{XMM-Newton} only. We again base our lag-energy spectra on Ark 564 \citep{Kara2013}, simulated for the two frequency ranges considered by \citet{Kara2016}: $[4-20]\times 10^{-5}$ Hz and $[3-11]\times 10^{-4}$ Hz. For the assumed exposure time, these two frequency ranges are both in the Gaussian error regime ($N=42$ and $N=208$). We see from Fig. 2 (third panel from the left) of \citet{Kara2016} that the lower frequency range is dominated by hard continuum lags. We therefore set our continuum lag parameters, $\phi_{\rm AB}$ and $\gamma$, to match the these observational data for this frequency range. The high frequency range (right hand panel of the same figure) is instead consistent with being dominated by reverberation lags, and so for this range we set $\phi_{\rm AB}=\gamma=0$ (following e.g. \citealt{Mastroserio2020}). The simulated data are shown in Fig \ref{fig:fitback}. The continuum lag dominated frequency range (panel b) and the reverberation dominated frequency range (panel c) match the observational data from Fig. 2 of \citet{Kara2016} very well. Note that the model for the high frequency lag-energy spectrum is set entirely by the spectral parameters, and it is therefore encouraging that it matches the observational data so well. For the simulation, we set the power spectrum of $A(t)$ to be $400$, and $25$ $({\rm fractional~rms})^2~{\rm Hz}^{-1}$ in lower and higher frequency range respectively. This is based on the observed power spectrum of NGC 4051 (see Fig 2 of \citealt{Vaughan2011}), which has a similar mass to Ark 564. We set the squared coherence of the two frequency ranges to $\gamma_c^2=$ $0.95$ and $0.8$. These parameters lead our simulation to have uncertainties consistent with the observed Ark 564 data.

\begin{table*}
\begin{center}
\renewcommand{\arraystretch}{1.5}
\begin{tabular}{ | l | c | c | c | c | c | } 
\hline
Parameter & Input & 1) X-rays only & 2) External Mass & 3) No soft X-rays & 4) Isotropic \\
          &   & Flat prior & Gaussian prior & Gaussian prior & Flat prior \\
\hline
\hline
$N_H$ ($10^{20}{\rm cm}^{-2}$) & 5 & $4.76^{+0.16}_{-0.18}$ & $4.81^{+0.16}_{-0.18}$ & $15.9^{+7.8}_{-12.4}$ & $4.73^{+0.15}_{-0.18}$ \\
\hline
$h$ ($R_g$) & 6 & $6.38^{+0.38}_{-0.37}$ & $6.27^{+0.36}_{-0.37}$ & $6.41^{+0.64}_{-0.69}$ & $5.96^{+0.21}_{-0.25}$\\
\hline
$a$  & 0.9 & $0.898^{+0.006}_{-0.006}$ & $0.897^{+0.005}_{-0.006}$ & $0.870^{+0.112}_{-0.052}$ & $0.896^{+0.006}_{-0.005}$\\
\hline
$i$ (degrees)  & 57 & $57.3^{+0.4}_{-0.6}$ & $57.2^{+0.4}_{-0.5}$ & $58.5^{+1.1}_{-1.4}$ & $57.4^{+0.4}_{-0.5}$ \\
\hline
$\Gamma$  & 2.45 & $2.44^{+0.01}_{ 0.004}$ & $2.44^{+0.01}_{ 0.001}$ & $2.47^{+0.02}_{-0.02}$ & $2.44^{+0.01}_{-0.003}$ \\
\hline
$D$ (Mpc)  & 100 & $118.3^{+14.9}_{-16.2}$ & $112.3^{+12.2}_{-12.0}$ & $100.8^{+15.8}_{-27.0}$ & $120.3^{+14.1}_{-17.7}$ \\
\hline
$A_{\rm fe}$ & 1 & $1.05^{+0.04}_{-0.06}$ & $1.04^{+0.04}_{-0.06}$ & $1.36^{+0.18}_{-0.37}$ & $1.05^{+0.03}_{-0.06}$ \\
\hline
$\log(n_e/{\rm cm}^{-3})$  & 17 & $17.0^{+0.1}_{-0.002}$ & $17.0^{+0.1}_{-0.01}$ & $17.0^{+0.2}_{-0.2}$ & $17.0^{+0.1}_{-0.006}$ \\
\hline
$(kT_e)_{\rm obs}$ (keV)  & 50 & $43.6^{+6.5}_{-13.7}$ & $43.3^{+7.2}_{-15.3}$ & $49.8^{+9.0}_{-15.1}$ & $44.7^{+5.9}_{-16.1}$ \\
 \hline
$1/\mathcal{B}$  & 1 & $1.30^{+0.29}_{-0.28}$ & $1.22^{+0.20}_{-0.31}$ & $1.16^{+0.36}_{-0.60}$ & $0.911^{+0.143}_{-0.129}$ \\
\hline
$M$ ($10^{6}~M_\odot$) & 3 & $3.13^{+0.29}_{-0.39}$ & $3.06^{+0.23}_{-0.34}$ & $3.24^{+0.43}_{-0.57}$ & $3.17^{+0.25}_{-0.36}$ \\
\hline
$h_d/r$ ($10^{-2}$) & 2 & $3.06^{+1.02}_{-3.01}$ & $2.63^{+0.86}_{-2.53}$ & $3.75^{+1.35}_{-3.78}$ & $3.40^{+1.55}_{-3.38}$ \\
\hline
$b_1$ ($10^{-2}$) & 0 & $19.0^{+4.8}_{-18.8}$ & $2.28^{+1.04}_{-2.34} \times 10^{-3}$ & $29.1^{+8.8}_{-30.0}$ & $0$ \\
\hline
$b_2$ ($10^{-2}$) & 0 & $12.7^{+11.3}_{-17.4}$ & $17.0^{+12.9}_{-21.0}$ & $28.1^{+34.2}_{-48.8}$ & $0$ \\
 \hline
$A$ ($10^{-4}$) & 2.2 & $2.04^{+0.06}_{-0.08}$ & $2.08^{+0.06}_{-0.08}$ & $2.21^{+0.16}_{-0.19}$ & $2.11^{+0.06}_{-0.07}$ \\
 \hline
 \hline
norm$_{\rm xill}$ ($10^{-5}$) & 6 & $6.61^{+0.85}_{-1.02}$ & $6.58^{+0.93}_{-1.00}$ & $6.80^{+1.22}_{-1.60}$ & $6.76^{+0.99}_{-0.98}$ \\
 \hline
 \hline
$\phi_{\rm AB}$ (rad) & -0.8 & $-0.837^{+0.118}_{-0.094}$ & $-0.820^{+0.116}_{-0.092}$ & $-0.974^{+0.181}_{-0.194}$ & $-0.818^{+0.107}_{-0.090}$ \\
 \hline
$\gamma$  & 0.3 & $0.301^{+0.012}_{-0.016}$ & $0.301^{+0.013}_{-0.015}$ & $0.303^{+0.014}_{-0.014}$ & $0.301^{+0.012}_{-0.015}$ \\
 \hline
 \hline
$H_0~3\sigma$ (${\rm km}~{\rm s}^{-1}{\rm Mpc}^{-1}$)  & 74.7 & $35-101$ & $41-101$ & $34-151$ & $34-95$ \\
 \hline
 \hline
 \hline
$\chi^2/$d.o.f.   &  & $2525.4/2606$ & $2525.4 /2606$ & $2034.6 /2071$ & $2535.4 / 2608$  \\
\hline
\hline
Steps used &  & 376,800 & 256,000 & 414,400 & 314,400   \\
Burn-in    &  & 519,200 (58\%) & 256,000 (50\%) & 219,200 (35\%) & 319,200 (50\%)  \\
\hline
\end{tabular}
\end{center}
\caption{Input parameters used for the simulation and posterior values with $1 \sigma$ uncertainties recovered from fitting the same model to the synthetic data. We trial four different model assumptions: 1) We fit with a flat prior on the mass, 2) We additionally include a Gaussian prior on the mass, 3) We ignore $E < 3$ keV X-rays but still employ a Gaussian mass prior, 4) We hardwire the corona to radiate isotropically except for the boost parameter ($b_1=b_2=0$). For all models, we fix $r_{\rm in}$ to the ISCO, $r_{\rm out}=2\times10^4$ and $z = 0.024917$. Whenever a mass prior is used, we assume an existing mass estimate of $M=[3.8\pm0.9] \times 10^6~M_\odot$. Some MCMC parameters are also listed. The steps quoted are total number of steps; e.g. 256,000 steps means 1000 steps per walker for 256 walkers. For each step, a new value of each free parameter is drawn.}
\label{tab:fitback}
\end{table*}

\subsection{Results}
\label{sec:results}

Using \textsc{xspec} v12.11.1 \citep{Arnaud1996}, we fit the input model to the synthetic data simultaneously over the \textit{XMM-Newton} lag-energy spectrum in two frequency ranges and the \textit{XMM-Newton}, \textit{NuSTAR} FPMA and \textit{NuSTAR} FPMB flux spectra. We begin our fit from the known input parameters, but note that it is in principle possible to find the global minimum $\chi^2$ from any starting point providing the model is correct. Here, we implicitly assume that the model provides the correct representation of the data in order to determine the statistical errors associated with data quality alone, whereas systematic errors associated with potentially incorrect or overly simplified model assumptions are discussed in Section \ref{sec:discussion}. For our baseline fit (Model 1), we consider an energy range of $0.5-10$ keV for \textit{XMM-Newton} and $3-50$ keV for \textit{NuSTAR}, and employ a flat prior on each parameter. The best fitting model is shown in Fig. \ref{fig:fitback} alongside the synthetic data, and the bottom panel shows the contributions to $\chi^2$ from different energy ranges. After minimising $\chi^2$, we run a Markov Chain Monte Carlo (MCMC) simulation within \textsc{xspec} to explore parameter space. We employ the Goodman-Weare \citep{Goodman2010} algorithm with 256 walkers. Table \ref{tab:fitback} displays the resulting peak posterior parameter values with $1\sigma$ uncertainties, as well as the chain length and burn-in period. Posterior distributions of 11 key model parameters are presented in Fig. \ref{fig:corner} and discussed in Appendix \ref{sec:mcmc}.

We additionally trial a number of fits with different model assumptions. Whereas Model 1 assumes no prior knowledge of the BH mass, we often in reality have an existing mass estimate from BLR reverberation mapping. To explore the prospects for using this information to improve our results, we run a new fit (Model 2) in which we include a Bayesian prior on the mass in the form of a Gaussian function. A typical precision of a BLR mass measurement is $30\%$ \citep{Peterson2004}. We therefore select a random variable from a Gaussian distribution with centroid $3\times 10^6~M_\odot$ and width $9\times 10^5~M_\odot$, to generate a synthetic BLR mass measurement of $M=[3.77 \pm 0.9] \times 10^6~M_\odot$. We include this fake BLR measurement by setting the centroid and width of our Gaussian mass prior to $3.77 \times 10^6~M_\odot$ and $9 \times 10^5~M_\odot$ respectively. The results are again quoted in Table \ref{tab:fitback}.

We see that the distance to our synthetic source can be recovered well. This measurement hinges on being able to constrain both $\xi$ and $n_e$ from the shape of the reflection spectrum. \citet{Garcia2016} show that $n_e$ influences the shape of both the soft excess and the iron line. Since our fits for Models 1 and 2 cover both of these features, they recover $n_e$ well. However, for various reasons discussed in Section \ref{sec:discussion}, the soft excess is less robust to modelling assumptions than the iron line is. Therefore, to determine if we can constrain the distance without fitting for the soft excess, we also fit Model 3. This model still employs a mass prior but ignores $E < 3$ keV for all spectra and lag-energy spectra. We see from Table \ref{tab:fitback} that this increases the uncertainty on $\log n_e$ by a factor of $\sim 4$, consequently increasing the uncertainty on $D$ by $\sim 40\%$. It is therefore desirable to consider soft X-rays but not necessarily vital, since constraints can be placed from the iron line alone.

Finally, we trial a model with a flat mass prior and $b_1=b_2=0$ hardwired (Model 4). This is to explore the importance of the static lamppost assumption. Since its assumptions are more restrictive, the Model 4 uncertainties will be smaller than those of Model 1. We see from Table \ref{tab:fitback} that this is indeed the case, but that the errors are not dramatically smaller. This indicates that, although the distance constraint of course has some sensitivity to the assumed angular emissivity of the corona, it does not critically hinge upon it.

\begin{figure*}
\centering
\includegraphics[width=18cm,trim=0.0cm 6.0cm 0.0cm 0.0cm,clip=true]{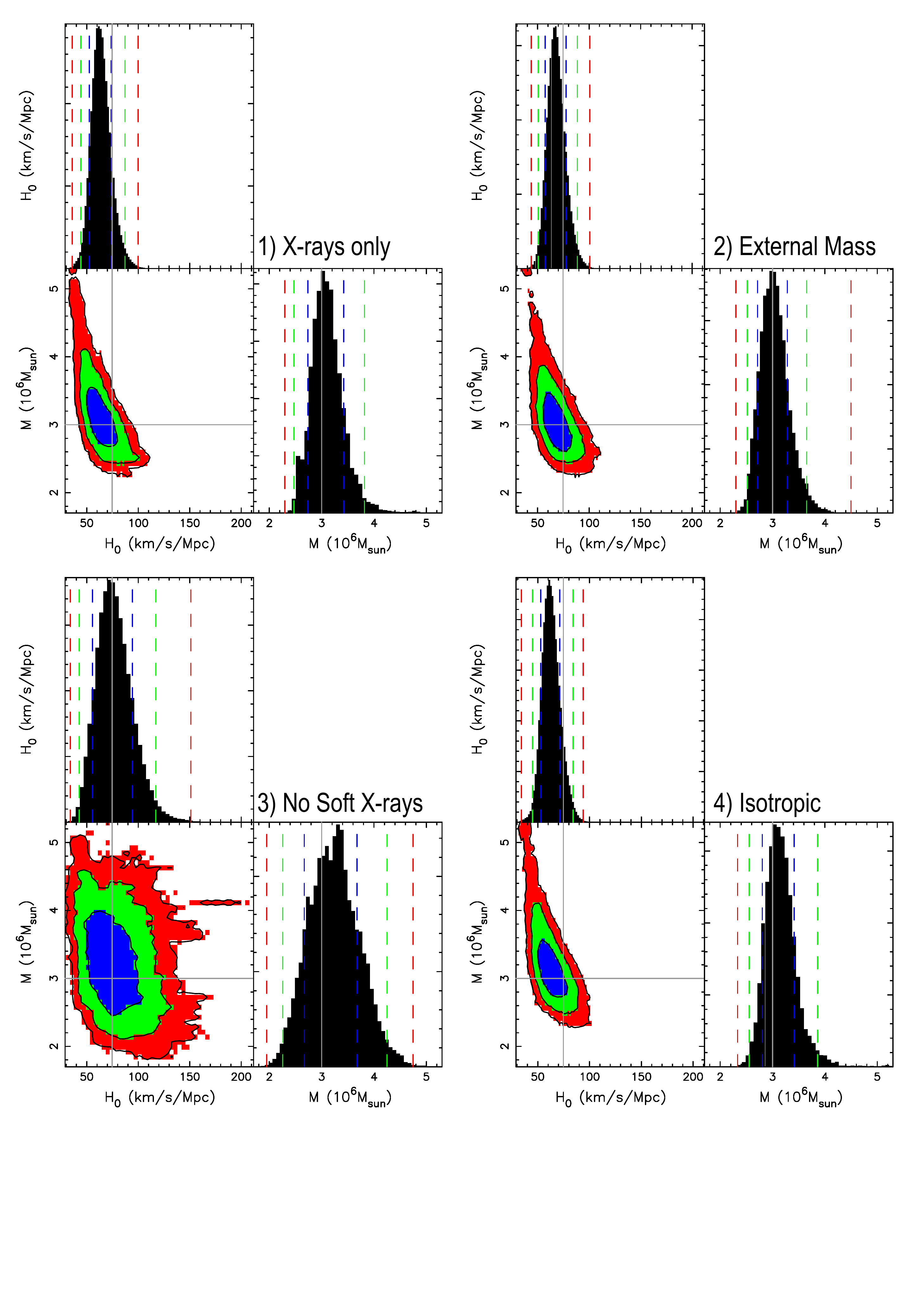}
\vspace{-3mm}
\caption{Posterior distributions of BH mass and Hubble constant extracted from our synthetic data with MCMC simulations. Each panel represents a different model fit to the same synthetic data: 1) Flat mass prior, 2) Gaussian mass prior ($M=[3.77\pm0.9] \times 10^6~M_\odot$), 3) same Gaussian prior but no $< 3$ keV X-rays, and 4) isotropic corona with a flat mass prior. $1$, $2$, and $3$ $\sigma$ confidence contours are shown in blue, green and red respectively, and the true values are depicted by solid grey lines. $H_0$ is calculated assuming a peculiar velocity drawn from a normal distribution with width $600~{\rm km}~{\rm s}^{-1}$.}
\label{fig:MvsH0}
\end{figure*}

We calculate a posterior distribution for $H_0$ from the MCMC values of $D$. For each step in the chain, we draw a value for the peculiar velocity of the host galaxy, $v_{\rm pec}$, from a Gaussian distribution with a mean of zero and standard deviation of $\bar{v}_{\rm pec} = 600 ~{\rm km}~{\rm s}^{-1}$ \citep[e.g.][]{Boruah2020} and calculate the Hubble constant as $H_0 = (zc+v_{\rm pec}) / D$. The addition of a peculiar velocity is required because the measured redshift $z$ is due to some combination of the expansion of the Universe and the host galaxy peculiar velocity. In the case of the $z \approx 0.02$ object we are simulating, peculiar velocity is therefore typically $\sim 10\%$ of the cosmological recession velocity. Peculiar velocity becomes less important as we consider higher redshift AGN, with the trade-off of course being that the count rate decreases with greater distance. The resulting posterior distributions of $H_0$ and $M$ are plotted in Fig. \ref{fig:MvsH0}. In Table \ref{tab:fitback}, we quote the $3~\sigma$ confidence range of $H_0$ for each of the four models trialled. As expected, the best constraint is for Model 2 ($41.0-101.0~{\rm km}~{\rm s}^{-1}{\rm Mpc}^{-1}$) and the worst is for Model 3 ($34.2-150.7~{\rm km}~{\rm s}^{-1}{\rm Mpc}^{-1}$).

We note that, in line with what has been demonstrated recently with observational data, the BH mass is well constrained by our analysis even when a flat prior is employed (Model 1). This is primarily because the reverberation delay depends on $h~R_g$ and $r_{\rm in}~R_g$, whereas the energy shifts to the reflected emission depend on $h$ and $r_{\rm in}$ \citep[see e.g.][]{Dauser2013}. This in turn is because the line profile from smaller $r$ is progressively broader, more skewed and more gravitationally redshifted. The observed line profile is the weighted sum of contributions from each radius. The disc inner radius governs the broadest line in the sum, and the corona height governs the emissivity weighting of the sum. Therefore simultaneous modelling of the flux spectrum and lag spectrum can constrain the mass. Moreover, we consider the lag-energy spectrum in two Fourier frequency ranges. This further constrains the mass and breaks the degeneracy (anti-correlation) with $h$ that occurs when only the lag-energy spectrum in one frequency range is considered. The degeneracy is broken because the reverberation lag depends on Fourier frequency: it is constant at the lowest frequencies, drops off above a break frequency due to the finite size of the reflector and is oscillatory at even higher frequencies due to phase-wrapping (see e.g. Fig 6 of \citealt{Bambi2021}). Therefore considering multiple frequency ranges exploits this frequency dependence in order to better constrain the BH mass. This point is discussed extensively in \citet{Ingram2019} and \citet{Mastroserio2020}.

\section{Discussion}
\label{sec:discussion}

We have shown from simulations that it is in principle possible to measure $H_0$ to a $3\sigma$ precision of $dH_0 \approx 30~{\rm km}~{\rm s}^{-1}~{\rm Mpc}^{-1}$ with X-ray reverberation mapping of archival X-ray data of a single AGN. Conducting this analysis for a sample of AGNs will reduce this uncertainty. Here we first discuss the sample size required to differentiate between distance ladder and CMB estimates of $H_0$ considering only statistical uncertainties. We then discuss the potential sources for systematic modelling error, and how they may be addressed.

\subsection{Statistical uncertainties}

For a sufficiently large sample of $K$ AGNs, the posterior distribution on $H_0$ becomes Gaussian with width $\langle dH_0 \rangle \approx dH_0 / \sqrt{K}$. Averaging different observations of the same AGN will reduce the uncertainty on $D$ but \textit{not} the uncertainty on the true cosmological redshift, which requires observations of multiple AGNs to average over different host galaxy peculiar velocities. The difference between the distance ladder and CMB estimates of $H_0$ is $\approx 6.5~{\rm km}~{\rm s}^{-1}~{\rm Mpc}^{-1}$. Therefore, to bring the $3\sigma$ statistical error down to this level will require $K\approx 22$ observations of comparable quality to the one we simulate here.

This estimate, however, does not include uncertainty on the absolute flux calibration of X-ray telescopes, which we can account for by adding it in quadrature to the statistical error. \textit{NuSTAR} has arguably the best absolute flux calibration of all the currently active X-ray telescopes. This is because the optics can be bypassed by collecting `stray light' from the source -- these are photons that enter the detectors without passing through the focusing optics. Without the optics, the instrument response is extremely simple to model. \citet{Madsen2017} were therefore able to measure the absolute flux of the Crab nebula by observing it off-axis in order to collect only stray light. By comparing the stray light observations with on-axis (focused) observations performed during the same year, they were able to calibrate the focused array. They estimate the systematic uncertainty on the response of the unfocused instrument to be $\sim 3\%$. Additionally, the intrinsic flux from the Crab can vary by up to $\sim 3\%$ from one observation to the next. Since the intrinsic flux may have varied between the un-focused and focused observations, the overall systematic error for the focused array can be conservatively estimated to be $\sim 6\%$. In future this second source of systematic error could be addressed by also monitoring the source with another observatory in order to quantify relative changes in flux between the focused and un-focused observations, but for the purposes of this discussion, we will consider a $6\%$ systematic error. Since distance relates to observed flux as $D \propto F_{\rm obs}^{-1/2}$ and $H_0 \propto 1/D$, $H_0$ relates to observed flux as $H_0 \propto F_{\rm obs}^{1/2}$. The systematic error on $H_0$ therefore relates to systematic error on $F_{\rm obs}$ as $dH_0/H_0=(1/2) dF_{\rm obs}/F_{\rm obs} \sim 3\%$. Taking this into account, it is still possible to achieve a $3 \sigma$ uncertainty of $\lesssim 6.5~{\rm km}~{\rm s}^{-1}~{\rm Mpc}^{-1}$ for a sample of $K \gtrsim 25$ sources. Note that if we are confident of the \textit{NuSTAR} absolute flux calibration, we can simply employ a free conversion factor in the fit to account for uncertainties in the \textit{XMM-Newton} calibration. In fact, since there are many joint \textit{XMM-Newton} and \textit{NuSTAR} observations of AGN, it should even be possible to constrain a conversion factor that can then be applied to \textit{XMM-Newton} observations with no joint \textit{NuSTAR} exposure.

Thus, if either of the two current best $H_0$ estimates is close to the true value, it is in principle possible to rule the other one out with $3\sigma$ confidence with a sample of $K\sim 25$ objects. If the true value is instead somewhere between the two best estimates, then a smaller uncertainty is required, and with it a larger sample. To date, approximately $\sim 25$ AGN have been reported to exhibit X-ray reverberation features in their lag spectra \citep[e.g.][]{Kara2016,Vincentelli2020}. Moreover, around the same number do not display a statistically significant iron line feature in their lag-energy spectra but are bright enough to achieve a reasonable signal to noise ratio. Our method is, in principle, applicable even to these objects, since an iron feature is not necessarily required to be significantly detected in the lag spectrum in order for BH mass to be constrained with the X-ray reverberation method (for instance, the lack of an iron feature places a strong upper limit on the reverberation lag). This was the case for Cygnus X-1 \citep{Mastroserio2019}.

The observation we simulate here is intended to be roughly representative of what is available in the archive. There are a number of AGNs with far more than $260$ ks of archival data, all of which can in principle be utilized to improve constraints. It is of course also possible to make new long exposure observations. It is also possible to additionally consider the \textit{NuSTAR} lag-energy spectrum, which requires the use of maximum likelihood techniques to account for Earth occultations \citep{Zoghbi2013,Wilkins2019}. In the near-mid future, X-ray observatories with a larger effective area will fly, providing better constraints; most notably \textit{the enhanced X-ray Timing and Polarimetry mission} (\textit{eXTP}; \citealt{eXTP2019}) and \textit{the Advanced Telescope for High ENergy Astrophysics} (\textit{ATHENA}; \citealt{Barret2020}). Providing good absolute flux calibration can be achieved, this will not only improve precision on distance measurements but it will also make it possible to conduct this analysis for AGN at slightly higher redshift for which peculiar velocity is less important.

\subsection{Systematic modelling uncertainties}

Our simulations quantify, for the first time, the statistical error on $H_0$ theoretically achievable from X-ray reverberation mapping of AGNs. It is encouraging, and not trivially apparent without simulations, that an uncertainty small enough to usefully address the $H_0$ tension is achievable with a relatively small sample of AGNs. It is therefore important to address potential sources of systematic error (i.e. incorrect model assumptions), which could in principle be far larger than the statistical error.

\subsubsection{The corona}

The most uncertain model assumption is perhaps the shape of the X-ray corona. Here we employ the lamppost model, which is the simplest model available to describe the observed spectral and timing signatures. The more compact the corona is in reality, the better the lamppost approximation is. The steep disc emissivity profile routinely inferred from the iron line profile provides indirect evidence that the corona is reasonably compact \citep[e.g.][]{Wilms2001,Fabian2009,Dauser2012}, and micro-lensing analyses constrain the hard X-ray half-light radius of strongly lensed quasars to be $\lesssim 10~R_g$ \citep[e.g.][]{Morgan2012}. Still, it is clear that more physically realistic coronal geometries (e.g. radially extended or vertically extended) must be investigated. The most basic extension to the lamppost model is the inclusion of a second lamppost source \citep{Chainakun2017}. The more sophisticated treatment of building up an extended corona from many small surface area elements has been implemented for the flux spectrum \citep{Dauser2013}, and shown to reproduce a similar disc emissivity profile (and therefore a similar spectrum) to the lamppost model. The time lags are more heavily influenced by the coronal geometry since they are additionally sensitive to the path length of rays from each surface area element, but previous calculations of timing properties for an extended corona \citep[e.g.][]{Wilkins2016} have been too computationally expensive to enable the model to be fit to data. The challenge in the near future is therefore to explore a range of physically realistic coronal geometries within a model framework efficient enough to enable model fitting.

Such a framework would also make it easier to address a key shortcoming of the current model: the coherence between energy bands is observed to drop off from $\sim$unity at high frequencies, but in our model it is always unity. This is essentially a result of assuming the corona to be point-like. In an extended corona, different regions can vary partially incoherently with one another, thus leading to non-unity coherence in the X-ray signal \citep[e.g.][]{Ingram2013}. A model that can predict the right coherence can then be fit to the modulus of the cross-spectrum as well as the lag-energy spectrum, which will provide stronger constraints \citep{Mastroserio2019}.

There are a number of routes to diagnose the coronal geometry from the data. For example, comparing the X-ray reverberation mass of an AGN with the BLR reverberation mass. These two measurements currently disagree for Mrk 335 \citep{Peterson2004,Mastroserio2020}, with the most likely cause being the lamppost assumption. Potentially stronger diagnostics can be provided by X-ray binaries with known distances and masses \citep[e.g.][]{Corral-Santana2016}. It is encouraging that the X-ray reverberation \citep{Mastroserio2019} and dynamical \citep{Miller-Jones2021} mass measurements of Cygnus X-1 now agree, and in future the new version of \textsc{reltrans} can be used to additionally measure a reverberation distance to compare with the parallax value. Another diagnostic that will soon be available upon the launch of the \textit{Imaging X-ray Polarimetry Explorer} (\textit{IXPE}; \citealt{Weisskopf2016}) will be X-ray polarization, which essentially measures the level of asymmetry and the importance of relativistic effects \citep[e.g.][]{Schnittman2010,Dovciak2011,Zhang2019,Chauvin2018}. In the current paper, we have included an empirical angular emissivity function to try and account for our ignorance of the true corona geometry. This treatment is inspired by the common use of an empirical radial emissivity profile in reflection studies \citep[e.g.][]{Dauser2010}. Allowing the angular emissivity parameters to be free in the fit therefore provides some reassurances that the measured parameters of interest are not driven critically by the restrictive assumptions that characterise the lamppost model. Our fit with the angular emissivity parameters free (Model 1) does indeed yield larger parameter uncertainties than our alternative model assuming isotropic emission (Model 4), but the increase is small. It is also encouraging that inferred distance is not correlated with the angular emissivity parameters ($b_1$, $b_2$ and \texttt{boost}) in our fits (see Fig. \ref{fig:corner}).

\subsubsection{Soft X-rays}

A number of sources of systematic error disproportionately affect the soft X-rays. One important example is absorption by partially ionized material around the AGN. Although the response of the absorbing gas to changes in the irradiating flux from the AGN does contribute its own time lag, this should only influence the lag-energy spectrum for Fourier frequencies lower than those of interest for our analysis \citep{Silva2016}. The influence of absorption on the soft X-ray region of the spectrum \citep[e.g.][]{LMiller2008,LMiller2010}, however, is potentially much more important and can influence the shape of the lag-energy spectrum \citep{Ingram2019}. It will therefore be prudent to select the AGN with the clearest view of the inner regions. The soft X-ray region of the reflection spectrum is also the most sensitive to modelling assumptions such as vertical disc structure \citep{Nayakshin2000,Done2007a,Rozanska2011,Vincent2016}. Moreover, X-ray reverberation models currently always assume that the time taken for photons to be re-processed and re-emitted in the disc atmosphere is very small compared with the light-crossing delays. This is a very good assumption for the fluorescence and scattering processes dominant for $E \gtrsim 3$ keV. However, the time taken for soft excess photons to be re-processed will be longer, since these photons undergo enough interactions to approximately thermalise. This thermalisation time is still expected to be small, but it is not yet clear if it is small enough to be neglected entirely. It is also important to note that the model we have explored here does not include a warm ($kT_e \sim 0.1$ keV), optically thick ($\tau \sim 10-40$) corona in addition to the hot corona, as is often invoked to explain the observed soft excess in AGNs \citep{Czerny1987,Middleton2009,Done2012,Petrucci2018,Ursini2020}. Models 1, 2 and 4 therefore effectively assume that the observed soft excess is dominated by reflection \citep[e.g.][]{Jiang2018,Garcia2019}. If in reality there is indeed a warm corona, its presence would make it more difficult to constrain the soft X-ray shape of the reflection spectrum \citep{Xu2021}. Since we believe our models to be most robust for $E \gtrsim 3$ keV, we ran an alternative fit ignoring $E < 3$ keV (Model 3), and found that the $3\sigma$ error on $H_0$ almost doubles ($dH_0 \approx 60~{\rm km}~{\rm s}^{-1}~{\rm Mpc}^{-1}$) compared to the other fits. It is therefore desirable to also include soft X-rays, but it is encouraging that constraints can even be achieved without. Ignoring soft X-rays entirely is an extreme measure. An alternative approach would be to keep the soft X-rays but trial a variety of different models, for instance with and without a warm corona. This approach would likely return uncertainties somewhere between the two extreme scenarios explored here.

\subsubsection{Disc physics}

There is also room to include more sophisticated disc physics in the model in future, although this should typically have a relatively small effect on overall measurements. For instance, the emergent reflection spectrum depends on the incident angle of irradiating photons, $\delta_i$. The \textsc{xillver} models hardwire this angle to $\delta_i=45^\circ$, whereas in reality $\delta_i$ should be a function of disc radius. We currently account for the leading-order effect of adjusting $\delta_i$ from $45^\circ$ by inputting into the \textsc{xillver} calculation an effective ionization parameter that differs from the calculated one \citep{Dauser2013,Ingram2019}. A more complete treatment would be to include $\delta_i$ as a parameter in the \textsc{xillver} grids. Here, we consider a disc with constant scale height $h_d/r$. This is an improvement on the common assumption of a flat disc ($h_d/r=0$), but in the \citet{Shakura1973} disc model the disc thickness is $\propto 1 - \sqrt{r_{\rm in} / r}$ and so increases steeply with $r$ close to the BH but is constant (i.e. $h_d/r \propto 1/r$) far from the BH. Accounting for this adjusts the disc emissivity profile and the light-crossing delays \citep{Taylor2018,Taylor2018b} due to e.g. shadowing of the inner disc for high inclination. We note however that the disc thickness is proportional to the Eddington ratio, and so $h_d/r=0$ is an increasingly good approximation for lower accretion rates.

Another effect not currently included in our model is returning radiation: reflected photons that return to the disc due to strong light bending and are reprocessed again. This increases the time delays since all returning photons have a longer path length than photons that only reflect once \citep{Wilkins2020}, and it also adjusts the emissivity profile (Dauser et al submitted). Returning radiation can be very important for $a \gtrsim 0.9$ and $h \lesssim 4$, and so any fits with parameters in this region should be treated with care if returning radiation is not included in the model. Finally, here we assume the disc to be planar, whereas a misalignment between the BH spin axis and rotation axis of the outer disc will warp the disc \citep{Bardeen1975,Liska2019}, which will again impact on the disc emissivity and light-crossing delays. Whether or not we expect AGN discs to be warped depends on whether or not the duration of a typical accretion event is shorter than the time it takes for the frame dragging effect to align the system \citep{King2006}.


\section{Conclusions}
\label{sec:conclusions}

We have shown that X-ray spectroscopy and reverberation mapping of a single bright Type 1 AGN can be used to constrain $H_0$ to a (statistical) $3~\sigma$ precision of $\sim \pm 30~{\rm km}~{\rm s}^{-1}{\rm Mpc}^{-1}$. It is therefore possible with a sample of $\sim 25$ such AGN to achieve the $\sim 6 ~{\rm km}~{\rm s}^{-1}{\rm Mpc}^{-1}$ precision required to address the current tension between distance ladder and CMB measurements. To this end, we have presented a new model from the \textsc{reltrans} package, called \textsc{rtdist}, which has distance to the source as a parameter instead of the disc ionization parameter. The model is designed to be used for AGN and X-ray binaries. X-ray binaries with existing mass and distance measurements can be used to test model assumptions. It is also of interest to measure the distance to X-ray binaries with the model in cases where it is not already known.
\textsc{rtdist}, as well as the model \textsc{simrtdist} that generates synthetic data from the model, will be made publicly available as user defined \textsc{xspec} models in a future release of the \textsc{reltrans} package.

\section*{Acknowledgements}

A. I. acknowledges support from the Royal Society. M. K. acknowledges support from an NWO (Nederlandse Organisatie voor Wetenschappelijk Onderzoek)
Spinoza grant. A. I. thanks K. Madsen for discussions about \textit{NuSTAR} flux calibration and R. Davies for discussions about galaxy peculiar velocity. We thank the anonymous referee for insightful comments.

\section*{Data Availability}

The \textsc{reltrans} package can be downloaded from \url{https://adingram.bitbucket.io/reltrans.html}. The updates described in this paper will be included in a future public release.
 



\bibliographystyle{mnras}
\bibliography{biblio} 




\appendix

\section{Energy shifts}
\label{sec:gfac}

The energy shift experienced by a photon going from point a to point b is
\begin{equation}
    g_{\rm ab} = \frac{ (p_b)^\mu (u_b)_\mu }{ (p_a)^\mu (u_a)_\mu },
\end{equation}
where $p^\mu$ and $u^\mu$ are 4-momentum and 4-velocity. The 4-velocity of a disc element is $u^\mu = u^t (1,0,0,\Omega)$, where $\Omega=1/(r^{3/2} \pm a)$ is angular velocity in units of radians per $t_g=R_g/c$ and
\begin{equation}
    u^t = \left( -\mathtt{g}_{tt} - 2 \mathtt{g}_{t\phi}\Omega - \mathtt{g}_{\phi\phi}\Omega^2 \right)^{-1/2},
\end{equation}
where $\mathtt{g}_{\mu\nu}$ is the metric. The Kerr metric components in the required sign and units convention are quoted in Appendix A of \cite{Ingram2015a}.

The energy shift from disc element to observer is therefore
\begin{equation}
    g_{\rm do} = \frac{-1}{p^\mu u_\mu} = \frac{\sqrt{-\mathtt{g}_{tt} - 2 \mathtt{g}_{t\phi}\Omega - \mathtt{g}_{\phi\phi}\Omega^2}}{-\mathtt{g}_{tt} - \mathtt{g}_{t\phi}( p^t \Omega + p^\phi ) - \mathtt{g}_{\phi\phi} p^\phi \Omega },
\end{equation}
where $p^t$ and $p^\phi$ are given by Equations (36) and (39) in \citet{Ingram2015a} with $\cos\theta = (h_d/r) / \sqrt{ (h_d/r)^2 + 1 }$ and $\theta_0=i$. The energy shift from source to disc element is
\begin{equation}
    g_{\rm sd} = \sqrt{ \frac{ (h^2-2h+a^2) / (h^2+a^2) }{ -\mathtt{g}_{tt} - 2 \mathtt{g}_{t\phi}\Omega - \mathtt{g}_{\phi\phi}\Omega^2 } }.
\end{equation}
In the limit $h_d/r=0$ both of these expressions reduce to their simpler equivalents quoted in I19. Even for $h_d/r=0.1$, the difference from the $h_d/r=0$ limit is $< 1\%$.



\section{MCMC}
\label{sec:mcmc}

For all four chains, we employ the Goodman-Weare algorithm \citep{Goodman2010,emcee2013} as implemented in \textsc{xspec} with 256 walkers. Each walker draws a new value of each free parameter for each step it takes, and the total number of steps we quote is always the number of steps per walker multiplied by the number of walkers. For each of the four chains, the Geweke convergence statistic \citep{Geweke1992} after the burn-in period is within $\pm 0.5$ for every parameter and for $\chi^2$, indicating good convergence.

\begin{figure*}
\centering
\includegraphics[width=\textwidth,trim=1.0cm 1.5cm 1.5cm 3.0cm,clip=true]{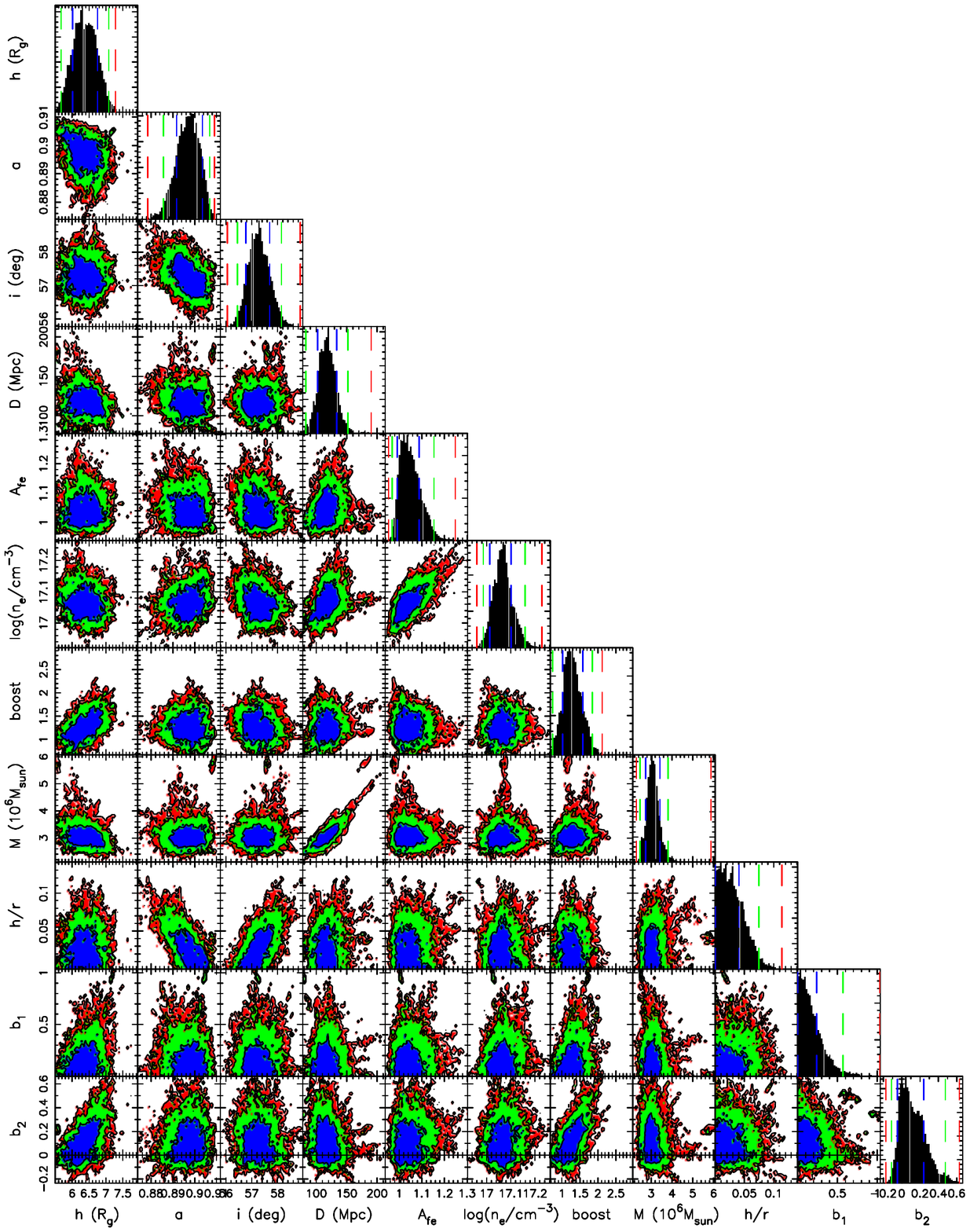}
\vspace{-5mm}
\caption{Model 1 posterior distributions of 11 model parameters extracted from our synthetic data with an MCMC simulation. $1$, $2$, and $3$ $\sigma$ confidence contours are shown in blue, green and red respectively.}
\label{fig:corner}
\end{figure*}

Fig. \ref{fig:corner} shows the posterior distributions of 11 key model parameters for the case of Model 1, for which a flat prior is employed for every parameter. Most parameters are only weakly correlated with one another. The strongest correlation is between $M$ and $D$. This is because the flux spectrum is only sensitive to $M/D$. This degeneracy is broken by the inclusion of the lag-energy spectrum, but since the signal to noise of the flux spectrum is so much higher than that of the lag-energy spectrum, the ratio $M/D$ is much better constrained than the individual parameters are. This correlation translates into a correlation between $H_0$ and $M$, but the correlation is significantly weakened by the dependence of $H_0$ on the host galaxy peculiar velocity (see Fig. \ref{fig:MvsH0}). $D$ is also weakly correlated with $\log n_e$. A correlation is expected because the flux spectrum is sensitive to $\xi \propto F_x / n_e$ and $F_x$ relates to the observed flux $F_{\rm obs}$ as $F_{\rm obs} \propto F_x / D^2$, therefore for a given value of $\xi$ we have $D^2 \propto n_e$. The correlation is weak because the flux spectrum appears to be almost as sensitive to $n_e$ as it is to $\xi$. $\log n_e$ is also correlated with $A_{\rm fe}$. This correlation is already known. For instance, \citet{Tomsick2018} found that a strongly super-solar iron abundance is required for Cygnus X-1 when using \textsc{xillver} (in which $n_e=10^{15}~{\rm cm}^{-3}$ is hardwired), but a more believable iron abundance of $\sim$unity results when using a high density reflection model.

The BH spin is weakly anti-correlated with the inclination angle, which is due to the trade-off between disc inner radius (here we fix inner radius to the ISCO) and inclination that goes into the shape of the iron line profile. The spin also anti-correlates with $h_d/r$, since they both influence the disc emissivity in the lamppost model. $h_d/r$ also correlates with inclination, since the disc opening angle governs the angle between the line of sight and disc normal for a given inclination angle. Finally, $b_2$ is correlated with $h$ and \texttt{boost} due to the effect they all have on the emissivity profile. It is encouraging for our purposes that $D$ is not correlated with many of the other model parameters.

\bsp	
\label{lastpage}
\end{document}